\documentclass[showpacs,preprintnumbers,amsmath,amssymb,onecolumn]{revtex4-2}

\usepackage[utf8]{inputenc}
\usepackage{hyperref}
\hypersetup{colorlinks=true,linkcolor=blue,citecolor=cyan}
\usepackage{array,multirow}

\usepackage{graphicx}
\usepackage{dcolumn}
\usepackage{bm}
\usepackage{color}
\usepackage{enumitem}
\usepackage{amsmath}
\usepackage{amssymb}

\newcommand{\bq}{\begin{equation}}
\newcommand{\eq}{\end{equation}}
\newcommand{\bqn}{\begin{eqnarray}}
\newcommand{\eqn}{\end{eqnarray}}

\usepackage{soul,xcolor}
\setstcolor{red}

\begin{document}

\title{A regular MOG black hole's impact on shadows and gravitational weak lensing in the presence of quintessence field}

\author{Ahmad Al-Badawi$^{1}$}
\email{ahmadbadawi@ahu.edu.jo}
\author{Sanjar Shaymatov$^{2,3,4,5}$}
\email{sanjar@astrin.uz}
\author{Mirzabek Alloqulov$^{6,5}$}
\email{malloqulov@gmail.com}
\author{Anzhong Wang$^{7}$}
\email{anzhong\_wang@baylor.edu}

\affiliation{$^1$Department of Physics, Al-Hussein Bin Talal University, P. O. Box: 20,
71111, Ma'an, Jordan}
\affiliation{$^2$Institute for Theoretical Physics and Cosmology, Zhejiang University of Technology, Hangzhou 310023, China}
\affiliation{$^3$Central Asian University, Milliy Bog Street 264, Tashkent 111221, Uzbekistan}
\affiliation{$^4$University of Tashkent for Applied Sciences, Str. Gavhar 1, Tashkent 100149, Uzbekistan}
\affiliation{$^5$Institute of Fundamental and Applied Research, National Research University TIIAME, Kori Niyoziy 39, Tashkent 100000, Uzbekistan}
\affiliation{$^6$New Uzbekistan University, Mustaqillik Ave. 54, Tashkent 100007, Uzbekistan}
\affiliation{$^{7}$GCAP-CASPER, Physics Department, Baylor University, Waco, TX 76798-7316, USA}

\date{\today}
\begin{abstract}
We investigate the impact of the modified gravity (MOG) field and the quintessence scalar field on horizon evolution, black hole (BH) shadow and the weak gravitational lensing around a  static spherically symmetric BH. We first begin to write the BH metric associated with the MOG parameter and quintessence scalar field. We then determine the BH shadow and obtain numerical solutions for the photon sphere and shadow radius. We show that the MOG ($\alpha$) and the quintessence ($c$) parameters have a significant impact on BH shadow and photon sphere. Based on the analysis, we further show that the combined effects of the MOG parameter and quintessential field can increase the values of BH shadow and photon sphere radii. We also obtain constraints on the BH parameters by applying the observational data of Sgr A$^{\star}$ and M87$^{\star}$.  Finally, we consider the weak deflection angle of BH within the context of the Gauss-Bonnet theorem (GBT) and show that the combined effects of the MOG and quintessence parameters do make the value of the deflection angle grow, referring to remarkable property being in well agreement with the physical meaning of both parameters that can maintain the strong gravitational field in the surrounding environment of BH.

\end{abstract}

\pacs{
}
\maketitle

\section{Introduction}

Einstein’s General Relativity (GR) has so far remained well-defined theory in which BH solutions to the field equations were predicted as simple mathematical models. This view point however changed as the existence of BHs in the universe has been confirmed by several modern observations in recent years, e.g., gravitational waves (GWs) \cite{Abbott16a,Abbott16b} and shadows of BHs by the Event Horizon Telescope (EHT) \cite{Akiyama19L1,Akiyama19L6}, both of which come into play the most important role 
in testing the nature of specetime geometry in the close surroundings of BHs. GR does however fail to explain the inevitable occurrence of singularities inside BHs, where it forecasts its own demise and loses its applicability. Hence, curvature singularity has long been regarded as a fundamental difficulty of GR as it cannot be explained by the theory itself. As a result of their efforts to find a new solution without singularity, numerous researchers proposed BH solutions in GR, usually known as regular BHs \cite{Bardeen68,Eyon,Hayward06,Moffat06MOG}. Since quantum theory gravity has not been well established, regular BHs can be used in BH physics and to explain geodesically complete spacetimes.
Also, GR would not be sufficient as potential explanation for dark matter, not directly detected, as well as the accelerated expansion of the universe. With this regard, GR remains an incomplete theory. To that it is widely believed that promising alternative theories are requisite to approach a fundamental understanding in relation to these objects. 

A well-established theory was proposed as an alternative to GR, which refers to scalar tensor vector gravity (STVG, i.e., Modified gravity (MOG)) resolving issues in relation to dark matter as well as dark energy (see details \cite{Moffat06MOG}). In STVG theory, a description of the expansion of the universe and the weakly interaction of massive particles, referred to as scalar and vector fields, was effectively described by its field equations solving to give solutions providing the MOG theory which can help to explain gravitational systems with their predictable nature. Afterwards, STVG theory opened up an active research field devoted to the study of its remarkable nature and attracted lots of attention in recent years. It should be emphasized that STVG solutions were widely applied to dark matter problems \cite{Moffat08:ApJ} as well as recent astrophysical observations \cite{Moffat06,Brownstein-Moffat07,Moffat16PLB}. There has since been an extensive analysis that considered various situations in the context of the MOG field theory; here we give some representative references \cite[see, e.g.,][]{Moffat07SS,Moffat09MNRASa,Moffat09MNRASb,Moffat21,Mureika16PLB,Pradhan19MOG,Rayimbaev20MOG,Qiao20MOG,Moffat19MNRAS,Moffat21JCAP,Turimov23PLB,Shaymatov23MRP,albadawi2023EPJC}.

From the astrophysical point of view it is particularly important to gain a deeper understanding about the nature of the existing fields in the BH surroundings and about their impacts on geodesics of massive and massless particles. Hence, such fields may alter the geodesics of massive and massless particles, i.g., observable properties such as the shadow, lensing, etc. In a realistic astrophysical scenario, it is increasingly important to consider the impact arising from dark energy in the surrounding environment of BHs as well as in large scales. Relying on the observations, it is widely believed that the expansion of the universe is getting accelerated as that of the vacuum energy, usually referred to as the cosmological constant $\Lambda$ in GR.  However, the quintessence scalar field has been well-defined and regarded as a promising alternative to the vacuum energy to examine the behaviour of dark energy~\cite{Peebles03,Wetterich88,Caldwell09}. First, Kiselev came up with the new solution including the quintessence scalar field, together with the state equation with $p = w_{q}\,\rho$ \cite{Kiselev2003aa}. Here, it should be noted that the equation of state parameter $w_q$ exhibits the quintessence field in the following range (-1;-1/3) as well as (-1;-2/3) \cite{Hellerman2001JHEP},  whereas $w_q=-1, \, -1/3$ refer to the vacuum energy with cosmological constant $\Lambda$ and other matter field, respectively. Following Kiselev \cite{Kiselev2003aa}, the quintessence and dark matter fields have since been considered in a variety of contexts~\cite{Li-Yang12,Shaymatov18a,Haroon19,Konoplya19plb,Hendi20,Jusufi19,Shaymatov21d,Rayimbaev-Shaymatov21a,Shaymatov21pdu}.

It is to be emphasized that the BH image was a fundamental question in the past decades. 
However, much progress was implemented in the observational studies of BHs with the help of recent modern astrophysical observations. To that, the recent triumphal discovery in relation to the detection of the first shadow image of supermassive BH referred to as the M87* galaxy~\cite{Akiyama19L1,Akiyama19L6} have confirmed not only the observational progress but also their existence in our universe. After that BH in astrophysics has now been taking center stage and its image has been tested not only by GR but also by modified theories of gravity. It is well defined that the gravitational lensing that the light can be strongly bent from its original path gives rise to the BH shadow as a result of reflection. It happens because the light cannot escape from BH's pull due to the strong deflection, thereby resulting in becoming a dark disc referred to as the BH shadow. Hence, BH shadow and the actual deflection angle are potentially crucial to help observers in testing the nature of geometry around the BH horizon. Note that the BH shadow as a disc was considered by Synge \cite{Synge66} and Luminet \cite{Luminet79} focusing on studying the light deflection in close vicinity of the Schwarzschild BH. Over the past years, BH shadows have widely been investigated and modeled theoretically; see for example    \cite{amarilla,Konoplya2019,Vagnozzi19,kumar2020,Afrin21a,abdujabbarov16,zhang,Atamurotov16EPJC,Konoplya19PRD,Atamurotov21JCAP,Mustafa22CPC,Tsukamoto18,Eslam-Panah20,Olmo23,Asukula23}. In recent years, BH shadows have been considered in the context of MOG theory \cite{Moffat15EPJC,Moffat15EPJCb,Moffat20PRD} addressing the shadows of non-rotating and rotating BHs. Note that a faraway observer cannot distinguish a BH geometry from other compact objects such as scalar boson and Proca stars being alternative to BHs. For the clear distinction, the shadow can be used as a viable test to explore their geometry. To this, BH shadow has been investigated very successful in determining their geometry and insights of scalar boson and Proca stars through the analytical fitting models using numerical solutions ~\cite{Rosa22,Rosa23}, and of regular BHs and geonic horizonless compact objects \cite{Olmo23}. Also, observational properties of relativistic fluid spheres with thin accretion disks have been considered with an extensive analysis \cite{Rosa23b}. Recent EHT observations are very important in studying BH shadows and helping to have restrictions on some theoretical models \cite{Hendi23}.

Another fundamental phenomenon is the gravitational lensing that occurs because of the light deviation under the strong gravitational field of massive central object, i.e., the light ray is bent from its original path due to the deflection. It is well established that the deflection can widely describe the gravitational lensing in GR. Hence, we further focus on studying the deflection angle around BH which not only provides the primary source of information pertaining gravity in the strong field regime and but also exhibits some departures of the geometry of compact objects. 
It is worth noting that GR was successfully tested first by the gravitational lensing effects to provide some evidences regarding to a BH geometry \cite{Eddington1919GL}. With this in view, the gravitational lensing has since widely been used as a key element to provide convincing comparisons and explanation for arbitrary theories of gravity. In this respect, a large amount of work has been implemented extensively to study the gravitational lensing in recent years \cite{Bisnovatyi-Kogan2010a,Tsupko12,Cunha20a,Atamurotov21PFDM,Javed22,Jafarzade21a,Atamurotov22,Atamurotov21galaxy,Atamurotov2022,Alloqulov_2023,albadawi2024shadows,10.1088/1674-1137/ad1677}. Later on, it was also extended to the modified theories of gravity \cite{Ovgun19WG,Rahvar19MNRAS,Izmailov19MNRAS}. The gravitational lensing has also been approached from different methods, e.g.,  one was proposed by Gibbons and Werner \cite{Gibbons08CQG,Werner12GBT}, referred to as the Gauss-Bonnet theorem (GBT). We further consider GBT on studying the weak deflection angle around the  MOG BH with quintessence field. 

In the present work, we consider a  MOG BH surrounded by quintessence field, as described by the line element in the next section. This geometry was proposed as a well-defined alternative to GR. We further aim to investigate the combined impact of the MOG and quintessence fields on horizons evolution, shadow, and the weak deflection angle for a static and spherically symmetric  MOG BH in the presence of quintessence field. 

This paper is organized as follows: In Sec. \ref{Sec:II} we briefly describe the metric of a static BH in the STVG theory with quintessence field. Furthermore, we obtain numerical results and plots for the horizons. In Sec. \ref{Sec:III}, we analyse the BH shadow and we study the impact of the model parameters on the photon sphere and shadow radius. In Sec. \ref{Sec:IV}, we study the weak deflection angle around the BH. Conclusions and final remarks are given in Sec. \ref{Sec:con}. We use a system of units in which $G_{\rm{N}}=c=1$ throughout the manuscript.

\section{ Regular MOG BH surrounded by quintessence field}\label{Sec:II}

\subsection{ A brief review of the spacetime}

Here, we consider a static and spherically symmetric BH in the presence of quintessence field in the STVG theory, which is given by  \cite{Moffat06MOG,Moffat15EPJC,Moffat21:EPJC,Kiselev2003aa}
\begin{equation}
ds^{2}=-f\left( r\right) dt^{2}+f^{-1}(r)dr^{2}+r^{2}\left( d\theta
^{2}+\sin ^{2}\theta d\phi ^{2}\right)\, ,   \label{M1}
\end{equation}%
where 
\begin{equation}
f\left( r\right) =1-\frac{2\left( 1+\alpha \right) Mr^{2}}{\left(
r^{2}+\alpha \left( 1+\alpha \right) M^{2}\right) ^{3/2}}+\frac{\alpha
\left( 1+\alpha \right) M^{2}r^{2}}{\left( r^{2}+\alpha \left( 1+\alpha
\right) M^{2}\right) ^{2}}-\frac{c\left( 1+\alpha \right) }{r^{3w_{q}+1}}\, , \label{mfunction}
\end{equation}
in which, $M$ is the mass of the gravitating object, $c$ is a normalization factor referred to as the quintessential field parameter that depicts the quintessence field's intensity and $w_{q}$ is the state parameter of the quintessence. The allowed values for $w_{q}$ are given in the range $-1< w_{q}<
-1/3$, as mentioned previously \cite{Kiselev2003aa,Hellerman2001JHEP}. We will refer to BH metric described by Eq.~(\ref{M1}) as a regular MOG BH with quintessence filed. We first examine the behavior of the MOG BH with quintessence solution, i.e., we explore the curvature scalars. The Ricci scalar for the given BH gemetry is defined by \begin{eqnarray}
R&=&\frac{6M^{3}\alpha \left( 1+\alpha \right) ^{2}(-r^{4}+2M^{3}\alpha
^{2}\left( 1+\alpha \right) (2M\left( 1+\alpha \right) -\sqrt{r^{2}+\alpha
\left( 1+\alpha \right) M^{2}})}{\left( r^{2}+\alpha \left( 1+\alpha \right)
M^{2}\right) ^{9/2}}\nonumber\\
&+&\frac{6M^{4}r^{2}\alpha ^{2}\left( 1+\alpha \right) ^{2}\left( 3M\left(
1+\alpha \right) +2\sqrt{r^{2}+\alpha \left( 1+\alpha \right) M^{2}}\right) 
}{\left( r^{2}+\alpha \left( 1+\alpha \right) M^{2}\right) ^{9/2}}+\frac{%
3cw_{q}\left( 3w_{q}-1\right) \left( 1+\alpha \right) }{r^{3\left(
w_{q}+1\right) }}\, . 
\end{eqnarray}
\begin{figure}
    \centering
{{\includegraphics[width=8.5cm]{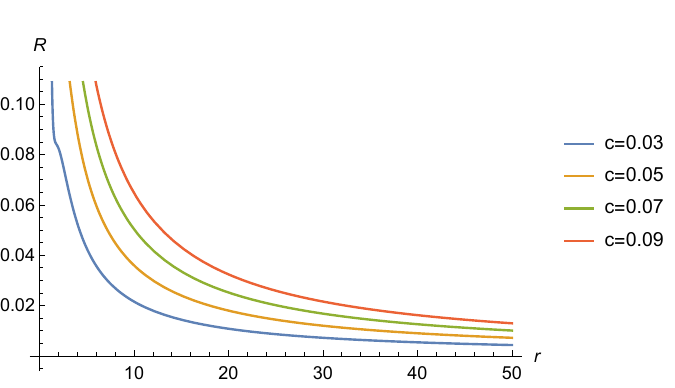} }}\qquad
    {{\includegraphics[width=7cm]{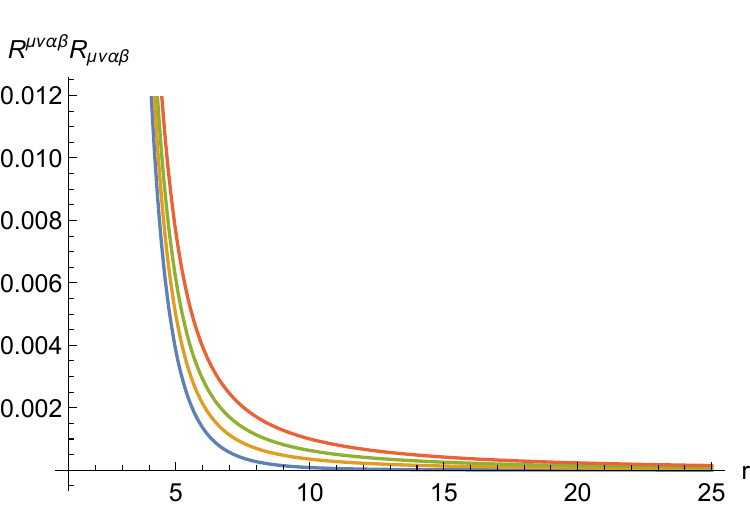}}}
    \caption{Variation of Ricci and Kretschmann scalars vs. $r$ for several  values of $c$. Here ($M=1$, $\alpha=0.2$ and $w_{q}=-2/3)$.}
    \label{figaaa1}
\end{figure}
\begin{figure}
    \centering
{{\includegraphics[width=8.5cm]{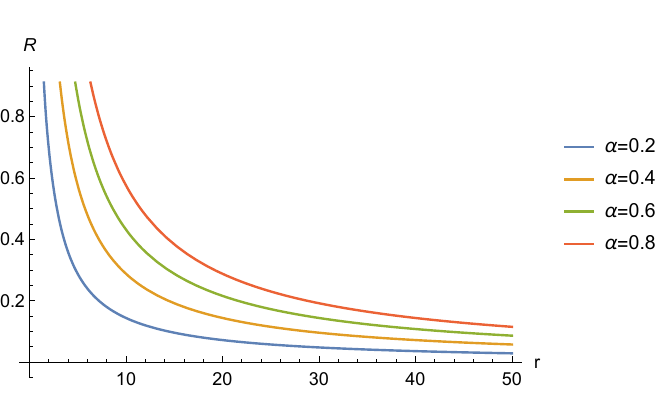} }}\qquad
    {{\includegraphics[width=7cm]{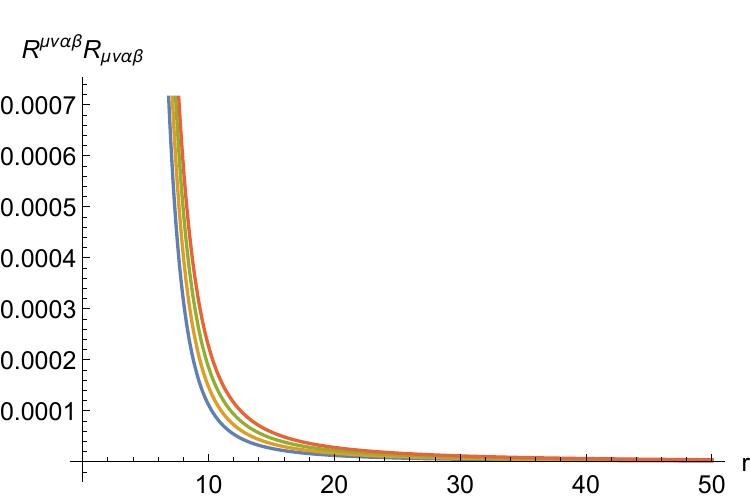}}}
    \caption{Variation of Ricci and Kretschmann scalars vs. $r$ for several  values of $\alpha$. Here ($M=1$, $c=0.02$ and $w_{q}=-2/3)$.}
    \label{figaaa2}
\end{figure}
From the above equation the Ricci scalar exhibits a physical singularity at $r = 0$ in the limit of $\alpha=0$ and $c= 0$. Figs. \ref{figaaa1} and \ref{figaaa2} provide a comprehensive analysis of the properties of MOG BHQ spacetime in terms of scalar invariants. Clearly, both scalar invariants follow the same behaviors and are positive-definite. Figs. \ref{figaaa1} and \ref{figaaa2} show that the MOG and quintessence parameters influence the variation of the Ricci scalar and the Kretshmann effectively when increasing $c$ and $\alpha$, thus resulting in increasing both scalars slowly. Furthermore, we observe that for very large values of $r$, all of the scalars approach to zero.  \begin{figure}
    \centering
    \includegraphics{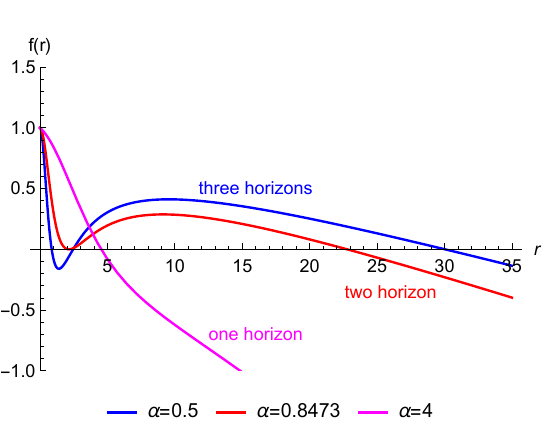}
    \caption{Metric function (\ref{mfunction}) of MOG BHQ for different values of the MOG
parameter $\alpha$}
    \label{figa1}
\end{figure}

\begin{figure}
    \centering
{{\includegraphics[width=7.5cm]{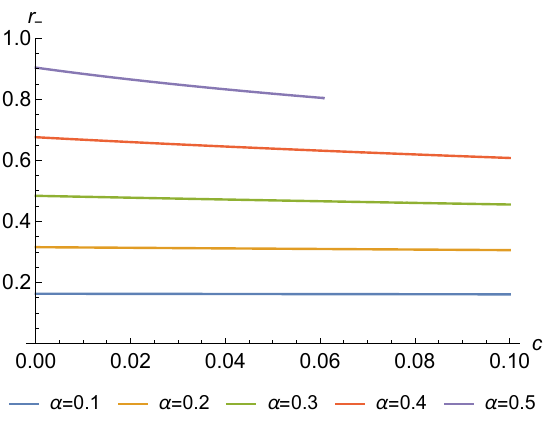} }}\qquad
    {{\includegraphics[width=7.5cm]{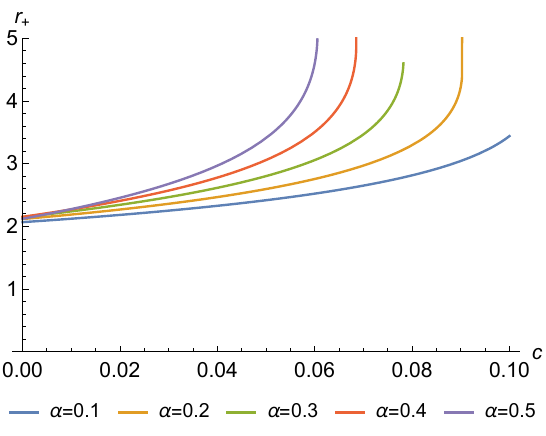}}}\qquad
    {{\includegraphics[width=7.5cm]{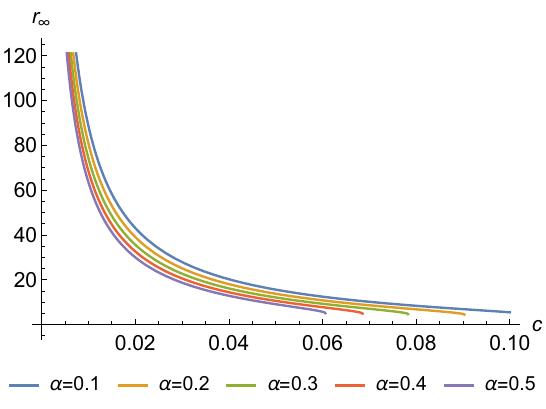}}}
    \caption{Plot of the three horizons vs. $c$ with different values of  $\alpha$. Here, $w_{q}=-2/3$. 
}
    \label{figa5}
    \end{figure}

It is well-defined that $f(r)=0$ can solve to give BH horizon, which is given by   
\begin{eqnarray}
\Big( 1-c\left( 1+\alpha \right) r\Big) \Big( r^{2}+\alpha \left(
1+\alpha \right) M^{2}\Big)^{2}&+&\alpha \left( 1+\alpha \right)
M^{2}r^{2}\nonumber\\&-& 2\left( 1+\alpha \right) Mr^{2}\sqrt{r^{2}+\alpha \left( 1+\alpha
\right) M^{2}}=0\, .  \label{hor2}
\end{eqnarray}%
The above equation has there roots that can represent Cauchy radius $r_{-}$, event horizon $r_{+}$ and cosmological horizon $r_{\infty}$. However, we further explore the horizon equation (Eq. (\ref{hor2})) numerically. The numerical results of these three horizons are tabulated in Table \ref{tabaa2}, in which we consider the role of the quintessence state parameter $w_{q}$ and the MOG parameter $\alpha$ on BH horizons. It can be seen from Table \ref{tabaa2}, the cosmological horizon is very sensitive with the effect of $w_{q}$, thus taking large values. 
\begin{table}
\begin{tabular}{|c|c|c c c| c c c|c c c|c c c|}   \hline

&   \multicolumn{13}{c|} {$w_q=-2/3$} \\ \hline

 & $\alpha =0$ & \multicolumn{3}{|c|}
 {$\alpha=0.1$}   &  \multicolumn{3}{|c|}
 {$\alpha=0.2$}   &  \multicolumn{3}{|c|}
 {$\alpha=0.3$} &     \multicolumn{3}{|c|}
 {$\alpha=0.4$}   \\ \hline
$c$ & $0$ & $0.01$ & $0.02$ & $0.03$ & $0.01$ & $0.02$ & $0.03$ & $0.01$ & $%
0.02$ & $0.03$ & $0.01$ & $0.02$ & $0.03$ \\ \hline
$r_{-}$ & x & $0.1626$ & $0.1624$ & $0.1622$ & $0.314$ & $0.313$ & $%
0.312$ & $0.480$ & $0.477$ & $0.474$ & $0.667$ & $0.659$ & $%
0.652$ \\ 
$r_{+}$ & $2$ & $2.12$ & $2.18$ & $2.25$ & $2.18$ & $2.26$ & $
2.35$ & $2.23$ & $2.34$ & $2.46$ & $2.27$ & $2.40$ & $2.57$
\\ 
$r_{\infty}$ & x & $88.65$ & $43.13$ & $27.92$ & $80.86$ & $39.11$ & $%
25.13$ & $74.23$ & $35.67$ & $22.73$ & $68.51$ & $32.67$ & $20.61$\\
\hline
&   \multicolumn{13}{c|} {$w_q=-4/9$}\\ \hline
$r_{-}$ & x & $0.1621$ & $0.1614$ & $0.1608$ & $0.3132$ & $0.3111$ & $%
0.3089$ & $0.4787$ & $0.4737$ & $0.4688$ & $0.6651$ & $0.6549$ & $0.6451$ \\ 
$r_{+}$ & $2$ & $2.0998$ & $2.1339$ & $2.1693$ & $2.1588$ & $2.2020$ & $%
2.2471$ & $2.2000$ & $2.2548$ & $2.3121$ & $2.2195$ & $2.2893$ & $2.3621$ \\ 
$r_{\infty}$ & x & $751308$ & $93907$ & $27819$ & $578697$ & $72330$ & $21426$ & $%
455158$ & $56888$ & $16850$ & $364423$ & $45545$ & $13489$\\ \hline
     \end{tabular}
     \caption{Numerical results for the Cauchy radius, $r_{-}$,  event horizon $r_{+}$
and cosmological horizon $r_{\infty}$ of the MOG BH surrounded by quintessence field. }
     \label{tabaa2}
\end{table}
Also, we illustrate the impact of the MOG parameter $\alpha $ and quintessence parameter $c$ on BH horizons in Fig. \ref{figa5}. As can be observed from Fig.~\ref{figa5}, both the Cauchy and event horizons increase, while the cosmological horizon decreases as a consequence of the increase in the value of MOG parameter $\alpha$. A rise in the quintessence parameter $c$ also leads to the decrease in the Cauchy radius and in the cosmological horizon, while to an increase in the event horizon. It is worth noting that both parameters $\alpha$ and $c$ have a significant impact on the cosmological horizon.

\section{Shadows of the  MOG BH with quintessence field }\label{Sec:III}

\begin{figure}
    \centering
    \includegraphics{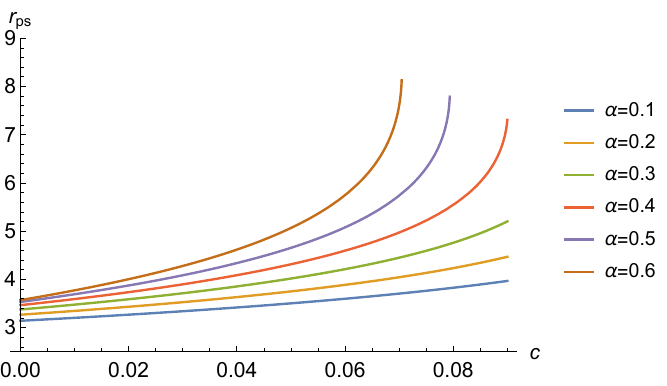}
    \caption{Dependence of the
photon sphere radius from the $\alpha$ and $c$ parameters.}
    \label{figa2}
\end{figure}

\begin{figure}
    \centering
    \includegraphics{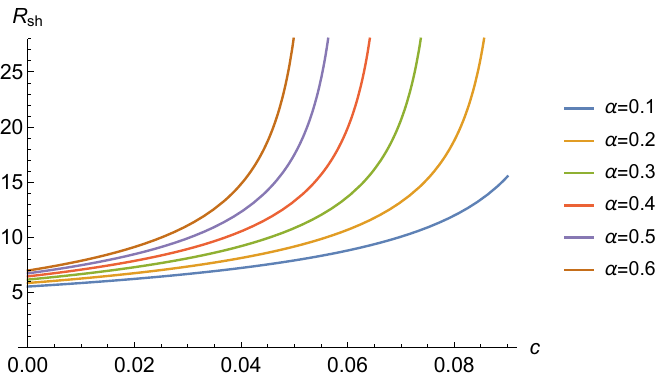}
    \caption{Variation of the shadow radius with the parameters  $\alpha$ and $c$.}
    \label{figa3}
\end{figure}

In this section, we study the shadow formation. To this we first consider the Hamilton-Jacobi equation for the system of test particle around the MOG BH with quintessence field, which is written as follows: 
\begin{equation}
\frac{\partial \mathcal{S}}{\partial \sigma }=-\frac{1}{2}g^{\mu \nu }\frac{%
\partial \mathcal{S}}{\partial x^{\mu }}\frac{\partial \mathcal{S}}{\partial
x^{\nu }},
\end{equation}%
where $\mathcal{S}$ is the Jacobi action. Consider the following Jacobi
action separable solution%
\begin{equation}
\mathcal{S}=-Et+\ell \phi +\mathcal{S}_{r}\left( r\right) +\mathcal{S}%
_{\theta }\left( \theta \right) ,
\end{equation}%
where the two constants $E$ (the energy) and $\ell$ (angular momentum) are given by\begin{equation}
E=\frac{dL}{d\dot {t}}=-\left( 1-\frac{2\left( 1+\alpha \right) Mr^{2}}{\left(
r^{2}+\alpha \left( 1+\alpha \right) M^{2}\right) ^{3/2}}+\frac{\alpha
\left( 1+\alpha \right) M^{2}r^{2}}{\left( r^{2}+\alpha \left( 1+\alpha
\right) M^{2}\right) ^{2}}-\frac{c\left( 1+\alpha \right) }{r^{3w_{q}+1}}\right) \dot{t}
\label{E1}
\end{equation}
\begin{equation}
\ell =\frac{dL}{d\dot {\phi}}=r^{2}\sin ^{2}\theta 
\dot {\phi }.  \label{An1}
\end{equation}
We can obtain the equations of
motion namely, 
\begin{equation}
\frac{dt}{d\sigma }=\frac{E}{f}, 
\end{equation}
\begin{equation}
\frac{d\phi }{d\sigma }=-\frac{\ell }{
r^{2}\sin ^{2}\theta },
\end{equation}
\begin{equation}
r^{2}\frac{dr}{d\sigma }=\pm \sqrt{\mathcal{R}\left( r\right) },  \label{R1}
\end{equation}
\begin{equation}
 r^{2}\frac{d\theta }{d\sigma }=\pm \sqrt{\Theta \left( \theta \right) } ,  
\end{equation}
where 
\begin{equation}
\mathcal{R}\left( r\right) =r^{4}E^{2}-\left( \mathcal{K}+\ell ^{2}\right)
r^{2}f,  \label{R2}
\end{equation}
\begin{equation}
\Theta \left( \theta \right) =\mathcal{K}-\ell ^{2}\cot \theta ,
\end{equation}
and $\mathcal{K}$ is the Carter separation constant.
Let us define the two quantities $\eta =\frac{\mathcal{K}}{E^{2}}$ and $%
\zeta =$ $\frac{\ell }{E}$ (stands for the impact parameters). Shadow casts can generally be obtained by using unstable null circular orbits. Therefore, Eq. (\ref{R1}) can be rewritten as 
\begin{equation}
\left( \frac{dr}{d\sigma }\right) ^{2}+V_{eff}=0,
\end{equation}%
where%
\begin{equation}
V_{eff}\left( r\right) =E^{2}\left[ \frac{\eta +\zeta ^{2}}{r^{2}}\left( 1-\frac{2\left( 1+\alpha \right) Mr^{2}}{\left(
r^{2}+\alpha \left( 1+\alpha \right) M^{2}\right) ^{3/2}}+\frac{\alpha
\left( 1+\alpha \right) M^{2}r^{2}}{\left( r^{2}+\alpha \left( 1+\alpha
\right) M^{2}\right) ^{2}}-\frac{c\left( 1+\alpha \right) }{r^{3w_{q}+1}}\right) -1\right] .
\end{equation}%
Following are the conditions for stable photons in circular orbits corresponding to the maximum effective potential:
\begin{equation}
V_{eff}\left( r\right) \left\vert _{r=r_{ps}}\right. =V_{eff}^{\prime
}\left( r\right) \left\vert _{r=r_{ps}}\right. =0, \label{v33}
\end{equation}
or
\begin{equation}
\mathcal{R}\left( r\right) \left\vert _{r=r_{ps}}\right. =\mathcal{R}%
_{eff}^{\prime }\left( r\right) \left\vert _{r=r_{ps}}\right. =0.  \label{R3}
\end{equation}%
where $r_{ps}$ is the radius of the unstable photon sphere orbit. We can obtain $r_{ps}$ using Eqs. (\ref{v33}) and (\ref{R3}) which is nothing but the solution of \begin{equation}
r_{ps}f'(r_{ps})-2f(r_{ps})=0,    
\end{equation} or 
\begin{eqnarray}
&&\frac{3}{2}cr^{-3w_{q}-1}\left( 1+w_{q}\right) \left( 1+\alpha \right) \left(
r^{2}+\alpha \left( 1+\alpha \right) M^{2}\right) ^{3}\nonumber\\&&-
\left( r^{6}+3M^{4}r^{2}\alpha ^{2}\left( 1+\alpha \right) ^{2}+M^{6}\alpha
^{3}\left( 1+\alpha \right) ^{3}\right)\nonumber\\&&+Mr^{4}\left( 1+\alpha \right) \left(
5M\alpha -3\sqrt{r^{2}+\alpha \left( 1+\alpha \right) M^{2}}\right)
=0. \label{rc12} 
\end{eqnarray}
Equation (\ref{R3}) yields the following result
\begin{equation}
\eta +\zeta ^{2}=\frac{4r^2_{ps}}{2f(r_{ps})+r_{ps}f'(r_{ps})}.
\end{equation}
Eq. (\ref{rc12}) is extremely difficult to solve analytically. As a result, we present numerical analysis and plots demonstrating the effects of the quintessence parameter $c$ and the MOG parameter $\alpha$ on the photon (circular) radius of mass particles and the radius of the BH shadow. Table \ref{taba1} displays the results of the photon sphere $r_{ps}$ and the BH shadow $R_{sh}$. For a constant value of $c$, our results show that $r_{ps}$ and $R_{sh}$ increase as $\alpha$ increases. When $\alpha$  is the same, increasing the parameter $c$ increases $r_{ps}$ and $R_{sh}$. In Figure \ref{figa2}, we show how the parameters $\alpha$ and $c$ influence $r_{ps}$ of massive particles surrounding a  MOG BHQ. It is clear that $r_{ps}$ rises as $\alpha$ and $c$ rise. Similarly, as shown in Fig. \ref{figa3}, the radius of the BH shadow increased as $\alpha$ and $c$ increased. 

\begin{figure}
    \centering
{{\includegraphics[width=7.5cm]{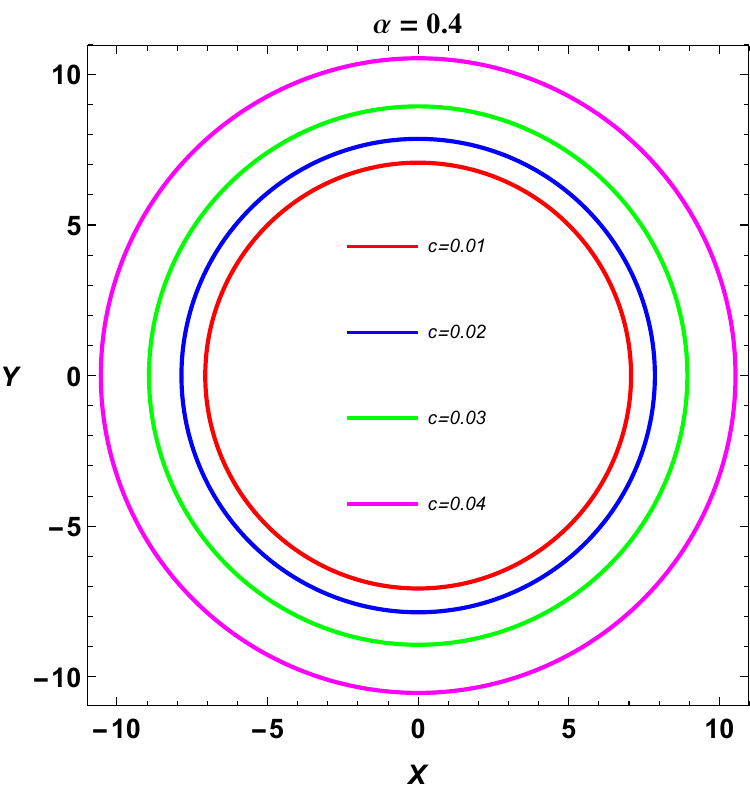} }}\qquad
    {{\includegraphics[width=7.5cm]{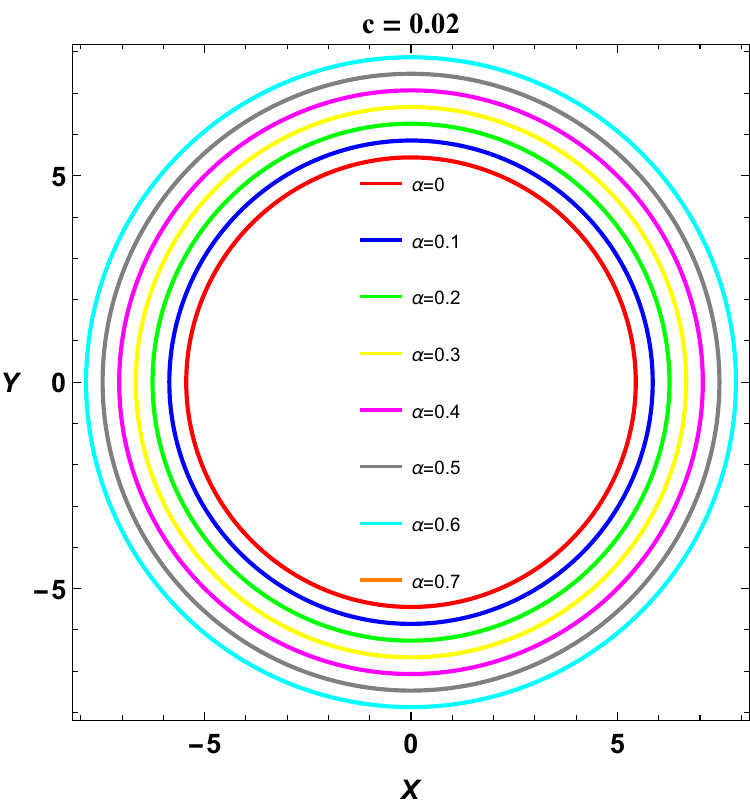}}}
    \caption{The shadow of the MOG BH surrounded by quintessence  field for different values of $c$ (left panel) and $\alpha$ (right panel). Note that we have set $w_q=-2/3$.}
    \label{figa4}
\end{figure}

Here, it should be noted that from the behaviour of the spacetime metric there exists the cosmological horizon which can be assumed to be located far away from the BH at a larger distance. Hence, we suppose that observer can be located at the cosmological radius. In the following, we will investigate how the size of the MOG  BHQ shadow radius, Rs, varies with $\alpha$ and $c$. We use the celestial coordinates $X$ and $Y$  \cite{vazquez} to locate the shadow for a more accurate representation. These coordinates are  \begin{equation}
X=\lim_{r_{0}\rightarrow \infty }\left( -r_{0}\sin \theta _{0}\left. \frac{
d\phi }{dr}\right\vert _{r_{0},\theta _{0}}\right) ,  \label{x11}
\end{equation}
\begin{equation}
Y=\lim_{r_{0}\rightarrow \infty }\left( r_{0}\left. \frac{d\theta }{dr}
\right\vert _{r_{0},\theta _{0}}\right) ,  \label{y11}
\end{equation}
 where $(r_{0},\theta_{0})$ are the position coordinates of the observer. Assuming the observer is on the equatorial hyperplane, Eqs. 
 (\ref{x11}) and (\ref{y11}) follow \begin{equation}
X^{2}+Y^{2}=\eta +\zeta ^{2}=R_{sh}^{2}.  \label{xy1}
\end{equation}  The size of the shadow can be analysed to determine the parameters including the parameters of the MOG BHQ. The variation of the size of the BH shadow has been depicted in Fig. \ref{figa4} .

\begin{table}   
     \begin{tabular}{|c|c c|c c|c c|c c|}  \hline
        & \multicolumn{2}{|c|}
 {$c=0.01$} & \multicolumn{2}{|c|}  {$c=0.02$} &  \multicolumn{2}{|c|}  {$c=0.03$} &  \multicolumn{2}{|c|}  {$c=0.04$}
   \\ \hline  
$\alpha$ & $r_{ps}/M$ & $R_{sh}/M$ & $r_{ps}/M$ & $R_{sh}/M$ & $r_{ps}/M$ & $R_{sh}/M$
& $r_{ps}/M$ & $R_{sh}/M$  \\ \hline
$0$ & 3.0464 & 5.4450 & 3.0958 & 5.7285 & 3.1487 & 6.0556 & 
3.2055 & 6.4387 \\ 
$0.1$ & 3.2030 & 5.8546 & 3.2688 & 6.2287 & 3.3401 & 6.6748 & 
3.4180 & 7.2189 \\ 
$0.2$ & 3.3467 & 6.2612 & 3.4328 & 6.7470 & 3.5276 & 7.3485 & 
3.6330 & 8.1203 \\ 
$0.3$ & 3.4770 & 6.6658 & 3.5882 & 7.2885 & 3.7128 & 8.0947 & 
3.8544 & 9.1964 \\ 
$0.4$ & 3.5928 & 7.0690 & 3.7354 & 7.8594 & 3.8979 & 8.9377 & 
4.0872 & 10.5358 \\ 
$0.5$ & 3.6924 & 7.4709 & 3.8742 & 8.4670 & 4.0854 & 9.9131 & 
4.3383 & 12.2997 \\ 
$0.6$ & 3.7728 & 7.8710 & 4.0044 & 9.1202 & 4.2786 & 11.0746 & 
4.6174 & 14.8262 \\ 
$0.7$ & 3.8286 & 8.2679 & 4.1254 & 9.8304 & 4.4820 & 12.5098 & 
4.9393 & 19.0008 \\ 
$0.8$ & 3.8494 & 8.6577 & 4.2358 & 10.612 & 4.7020 & 14.3736 & 
5.3269 & 28.4836%
\\ \hline
     \end{tabular}
     \caption{Numerical results for the $r_{ps}$ and  $R_{sh}$ of the MOG BHQ. Here, $w_q=-2/3$.}
     \label{taba1}
\end{table}


\begin{figure}
    \centering
    \includegraphics[width=0.5\textwidth]{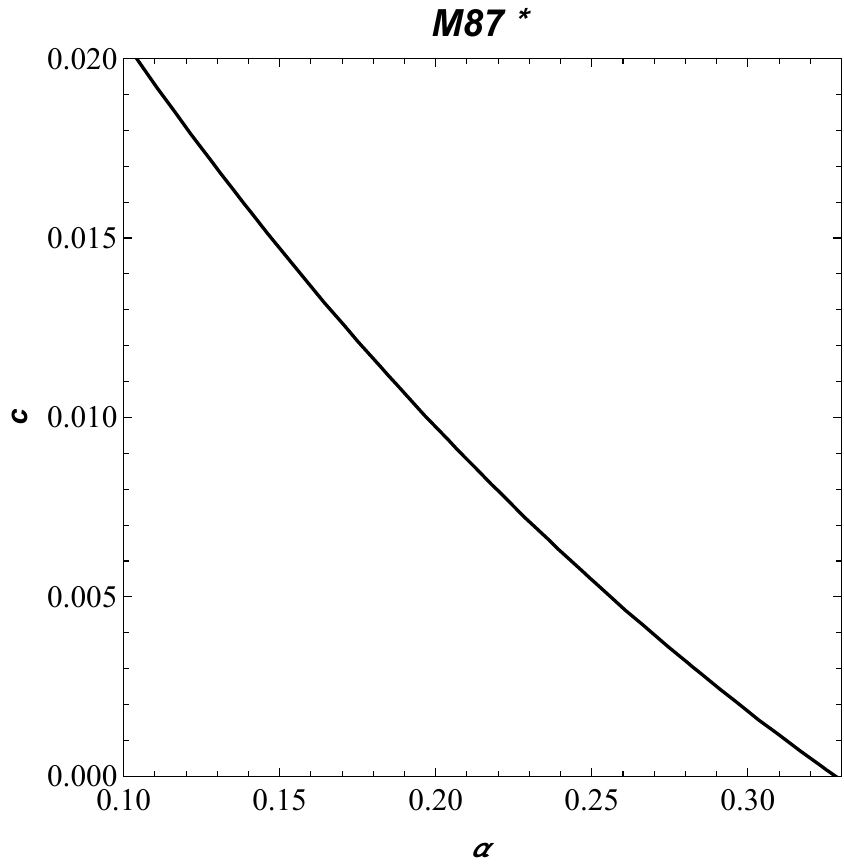}%
    \includegraphics[width=0.5\textwidth]{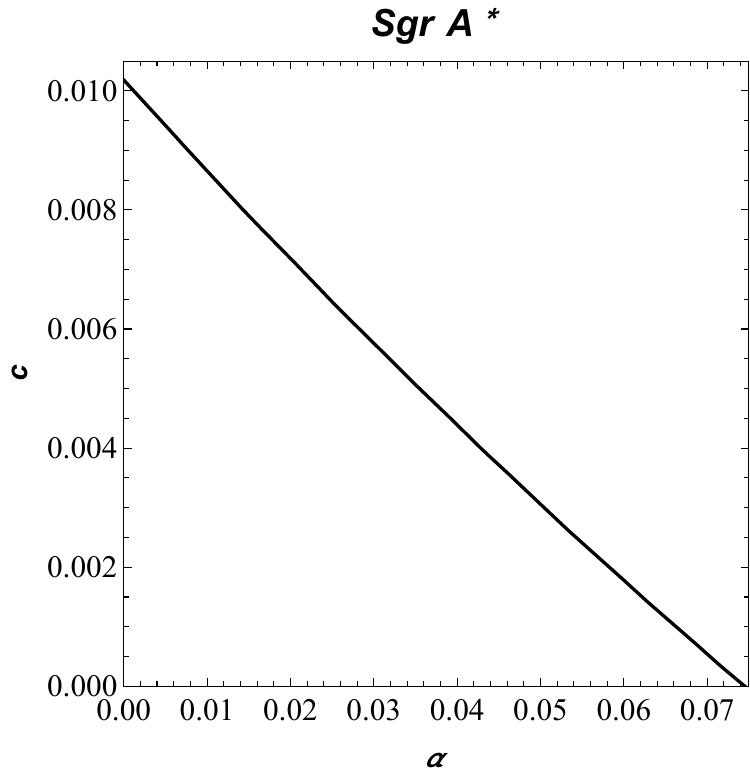}
    \caption{Constraint values of $\alpha$ and $c$ for M87$^{\star}$ and Sgr A$^{\star}$. Here, we have set $M=1$ and $w_q=-2/3$.}
    \label{fig:constraint}
\end{figure}

Now we turn to the application of theoretical results to the recent EHT observations, which can help us to have restrictions on the BH parameters. For our purpose we suppose that both Sgr A$^{\star}$ and M87$^{\star}$ can be regarded as static and spherically symmetric BHs in our model. As a matter of fact that they may not support our approach considered here. However, they satisfy well our assumption. In doing so, we aim to constrain the lower limits of both parameters $\alpha$ and $c$, relying on the EHT observational data in connection with the shadows with together the angular diameter of BH shadow $\theta$, the distance $D$ and the mass of the supermassive BHs Sgr A$^{\star}$ and M87$^{\star}$, e.g. $\theta_{M87^{\star}}=42 \pm 3 \mu as$, $D= 16.8 \pm 0.8 M pc$ between Earth and M87$^{\star}$ and $M_{M87^{\star}} = 6.5 \pm 0.7 \times 10^9 M_{\odot}$ for M87$^{\star}$ and $\theta_{Sgr A^{\star}}=48.7 \pm 7 \mu$,  $D=8277 \pm 9 \pm 33 pc$ and $M_{Sgr A^{\star}} = 4.297 \pm 0.013 \times 10^6 M_{\odot}$ for Sgr A$^{\star}$ (VLTI), respectively  \cite{Akiyama19L1,Akiyama19L6}. Taking these observational data into consideration, we determine the shadow diameter per unit mass of the BH as stated by the following expression
\begin{eqnarray}
    d_{sh}=\frac{D\theta}{M}\, .
\end{eqnarray}
Following $d_{sh}=2R_{sh}$, we further obtain the BH shadow's diameter. As a result, we define it as follows: $d^{M87^{\star}}_{sh}=(11 \pm 1.5)M$ for M87$^{\star}$ and $d^{Sgr^{\star}}_{sh}=(9.5 \pm 1.4)M$ for Sgr A$^{\star}$, respectively. Based on the observational EHT data, we explore constraints on both parameters $\alpha$ and $c$ numerically for Sgr A$^{\star}$ and M87$^{\star}$. By applying the observational data we then show the lower values of $\alpha$ and $c$ in Fig.~\ref{fig:constraint}. As can be observed from Fig.~\ref{fig:constraint}, one can expect low values of both parameters from observational data of Sgr A$^{\star}$ as compared to the one for M87$^{\star}$. However, it is worth noting that the observational data from the shadow of these two objects considered here are only facts to obtain the best-fit constraints on these parameters $\alpha$ and $c$.  

\begin{figure*}
\centering
 \includegraphics[width=0.5\textwidth]{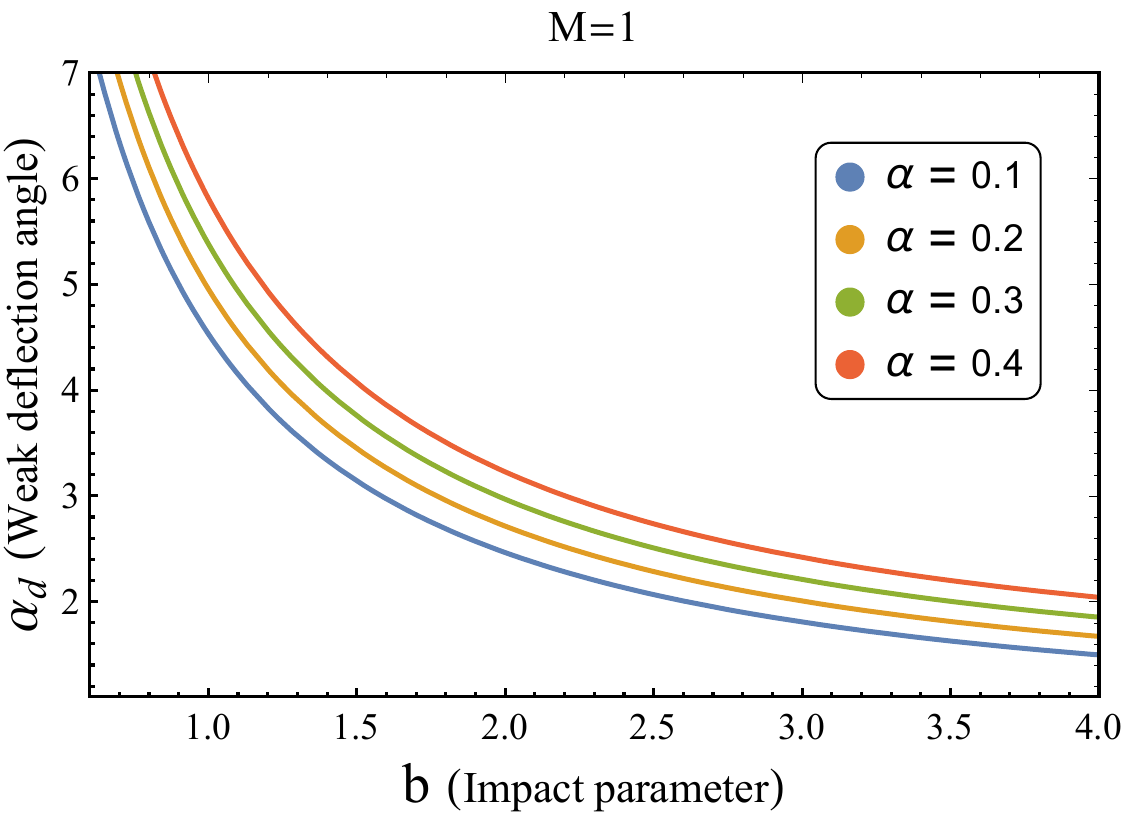}%
 \includegraphics[width=0.5\textwidth]{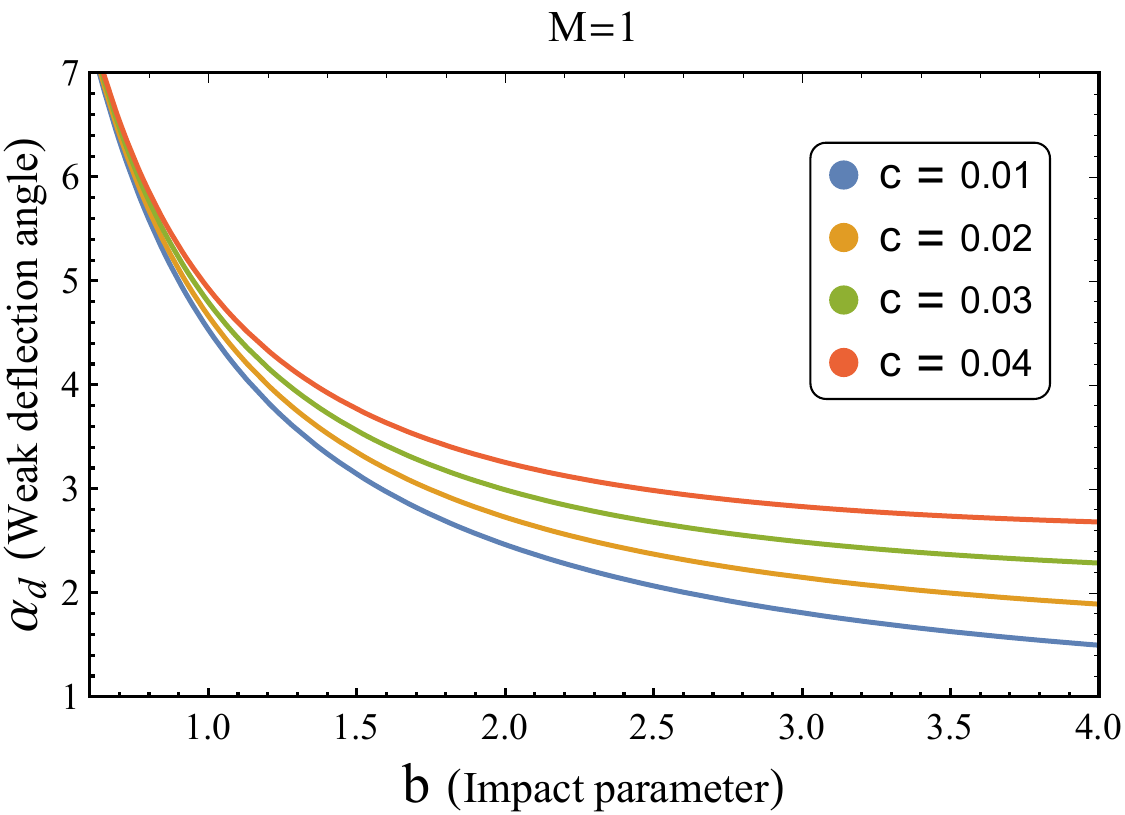}
 
  \caption{The deflection angle around the  MOG BH surrounded by quintessence for various combinations of MOG parameter $\alpha$ for fixed $c=0.01$ (left) and of quintessential parameter $c$ for fixed $\alpha=0.1$ (right column). Note that we have set $M=1$ and $\omega_q=-2/3$.}
\label{Fig:def_angle}
\end{figure*}

\section{Weak deflection angle around the  MOG BH surrounded by quintessence field}\label{Sec:IV}

Here, we turn to the study of the weak deflection angle around the  MOG BH surrounded by the quintessential field. To this end, we approach this issue from the prospective of the well-accepted Gauss-Bonnet theorem (GBT) developed by Gibbons and Werner as a new thought experiment helping one in evaluating the weak deflection angle of spherically symmetric BH spacetimes (see details in Refs.~\cite{Gibbons08CQG,Werner12GBT}). Afterwards, this thought experiment was also extended to various situations, e.g., the axisymmetric spacetimes \cite{Ono17GBT}, non-asymptotically flat spacetimes \cite{Ishihara16GBT,Ishihara16} as well BH spacetimes with a plasma medium \cite{Crisnejo18BGT}. There has since been an extensive analysis done as stated by the GBT method in recent years (see for example \cite{Li20,Jusufi18GBT,Ovgun18GBT,Ovgun19GBT,Arakida20GBT,Zhang21GBT,DCarvalho21,Kumar19,Fu21GBT,Mustafa22CPC}). 

In the following, we also consider the weak deflection angle of BH within the context of the GBT method. For that, we shall further restrict the motion to null geodesics for the photon, and we shall then consider the optical metric for the  MOG BH surrounded by quintessential field with the corresponding line element, which can be defined accordingly by 
\begin{eqnarray}
d\sigma^{2} = g_{kl}^{\mathrm {opt}}dx^{k}dx ^{l} =\frac{1}{f(r)}\left(\frac{d r^{2}}{f(r)}+r^2 d \phi^{2}\right)\, ,
\end{eqnarray}
with 
\begin{eqnarray}
f\left(r\right) &=& 1-\frac{2\left(1+\alpha \right) Mr^{2}}{\Big(
r^{2}+\alpha \left( 1+\alpha \right) M^{2}\Big) ^{3/2}}
+\frac{\alpha
\left( 1+\alpha \right) M^{2}r^{2}}{\Big( r^{2}+\alpha \left( 1+\alpha
\right) M^{2}\Big)^{2}}-\frac{c\left( 1+\alpha \right) }{r^{3w_{q}+1}}\, . \nonumber
\end{eqnarray}
The above mentioned optical metric helps us in determining the Gaussian curvature $K$ with the terms of radial coordinate, i.e., it is given with linear order of $M$ as follows:   
\begin{eqnarray}
KdS&=&
\frac{1+\alpha}{4 r^2 (1-(\alpha +1) c r)^{5/2}}\Big[(1+\alpha) c^2 r^3 \big((1+\alpha) c r-1\big)\nonumber\\&-& \big(5 (1+\alpha) c r \left(3 (1+\alpha\right) c r-4)+8\big)M\Big]\, ,
\end{eqnarray}
where we have set $\omega_q=-2/3$. 
In relation to the contribution stemming from the MOG field, which is given to the Gaussian curvature, one can write the following form for the geodesic curvature \cite{Gibbons08CQG,Crisnejo18BGT}
\begin{eqnarray}
\frac{d\sigma}{d\phi}\bigg|_{C_{R}}=
\left( \frac {r^2}{f(R)} \right)^{1/2}\, .
\end{eqnarray}
The above expression leads to the following form in the limiting case 
\begin{eqnarray}
\lim_{R\to\infty} \kappa_g\frac{d\sigma}{d\phi}\bigg|_{C_R}\approx 1\, .
\end{eqnarray}
Keeping the above expressions in mind together with the limiting case $R\to\infty$ and parametrization $r=b/\sin\phi$, we can further determine the deflection angle on the basis of the GBT method, which is rewritten as \cite{Gibbons08CQG}
\begin{eqnarray}
 \alpha_d &=&\int^{\pi+\alpha_d}_0 \left[\kappa_g\frac{d\sigma}{d\phi}\right]\bigg|_{C_R}d\phi-\pi \nonumber\\&=& -\lim_{R\to\infty}\int^\pi_0\int^{\infty}_{\frac{b}{\sin\phi}}K\,dS\, ,
\end{eqnarray}
where we have defined $b$ as the impact parameter. 
As a result, the deflection angle around the  MOG BH with the quintessence can be determined by the following approximate form
\begin{eqnarray}\label{Eq:def_angle}
\alpha_d \approx
\frac{( 1+\alpha) M \Big[4+5 \pi  (1+\alpha )c\, b \log (2 b)\Big]}{b}\, .
\end{eqnarray}
We will examine the impact of the MOG and quintessential fields on the deflection angle in the weak form as stated by the GBT method considered here. To be more informative and to understand more deeply, we show the dependence of the deflection angle $\alpha_d$ on the impact parameter $b$ for various combinations of the MOG and quintessential field parameters. As can be observed from Fig.~\ref{Fig:def_angle}, the deflection angle decreases as the impact parameter increases. It is however obvious that the parameters $\alpha$ and $c$ have the physical impact that can shift the deflection angle upward toward to its larger values. This is consistent with the physical meaning of both parameters as attractive gravitational charges, thereby manifesting the model for the MOG and quintessential fields and their profile on the weak deflection angle.  

\section{Conclusion}\label{Sec:con}

In this paper, we investigated the effect of the MOG and the quintessence scalar fields on the optical properties of static and spherically symmetric regular BH. In pursuit of this goal, we focused on studying the horizon evolution, BH shadows, the constraints on the BH parameters through the EHT observational data as well as the weak gravitational lensing. 

The first property of the spacetime of MOG BH surrounded by quintessence is the locations of three horizons. We obtained numerical solutions for the locations of the horizons represented in Table \ref{tabaa2} and Fig. \ref{figa5}. It was found that that the quintessence state parameter $w_{q}$ significantly affects the cosmological horizon, reducing its radius significantly. We also showed that the MOG parameter $\alpha$ increases the Cauchy and event horizons, while it decreases the cosmological horizon. It was also shown that the decrease in the Cauchy radius and event horizon as well as a rapid fall in the cosmological horizon occur due to the increase in the quintessence parameter $c$. 

We further focused on studying the BH shadow and explored the photon sphere and the shadow radius numerically (Table \ref{taba1}). We demonstrated that the MOG $\alpha$ and quintessence field $c$ parameters have a significant impact on the BH shadow and photon sphere (Figs. \ref{figa2}-\ref{figa3}). Based on our findings, we demonstrated that the combined effects of the MOG and the quintessence field parameters can raise the values of BH shadow and photon sphere radii in comparison to the Schwarzschild case. Also, taking into account results regarding to BH shadow, we obtained constraints on the parameters $\alpha$ and $c$ as inferred from the EHT observational data for Sgr A$^{\star}$ and M87$^{\star}$. As a result, we showed that both parameters can take lower values as a consequence of observational data of Sgr A$^{\star}$ in contrast to the one for M87$^{\star}$. This is in well agreement with the fact of M87$^{\star}$ which is more massive and larger than Sgr A$^{\star}$.  

We also studied the impact of the MOG and quintessence fields together on the deflection angle and we showed that the deflection angle decreases as a consequence of the increase in the value of the impact parameter. The parameters both $\alpha$ and $c$ have similar effect, thereby resulting in increasing the deflection angle. However, the combined effects of both $\alpha$ and $c$ on the deflection angle become more powerful, as can be observed in Fig.~\ref{Fig:def_angle}. This behavior is in well agreement with the physical meaning of both $\alpha$ and $c$ as attractive gravitational charges, thus strengthening the gravity and manifesting the model more effective on the weak deflection angle. 

The theoretical studies and results can help to open up new avenues for observational testing of this kind of objects, as well as distinguishing between different geometries of compact objects. The research into applying similar investigation to rotating BHs might be conducted in the future.

\begin{acknowledgments}

The research is supported by the National Natural Science Foundation of China under Grant No. 11675143 and the National Key Research and Development Program of China under Grant No. 2020YFC2201503. M.A wishes to acknowledge the support from Research Grant F-FA-2021-432 of the Ministry of Higher Education, Science and Innovations of the Republic of Uzbekistan.

\end{acknowledgments}


\bibliography{gravreferences,Ref}

\end{document}